\documentclass[prc,aps,twocolumn,showpacs,preprintnumbers,nofootinbib,amsmath,amssymb]{revtex4}
\usepackage{graphicx}
\usepackage{dcolumn}
 \def\be{\begin{equation}}
 \def\ee{\end{equation}}
 \def\bea{\begin{eqnarray}}
 \def\eea{\end{eqnarray}}
 \def\bean{\begin{eqnarray*}}
 \def\eean{\end{eqnarray*}}

 \def\l{\left}
 \def\r{\right}

 \def\bm#1{\mbox{\boldmath$#1$}}
 
 \def\ksim{\mathrel{\rlap{\lower0.2em\hbox{$\sim$}}\raise0.2em\hbox{$<$}}}

\begin{document}
\title{Parton transport and hadronization from the dynamical quasiparticle point of view}

\author{W. Cassing$^a$ and E. L. Bratkovskaya$^b$}

\affiliation{ \phantom{a}
 $^a$Institut f\"{u}r Theoretische Physik,
      Universit\"{a}t Giessen, 35392 Giessen, Germany \\
$^b$Frankfurt Institute for Advanced Studies,
      JWG Universit\"{a}t Frankfurt, 60438 Frankfurt am Main, Germany}

\begin{abstract}
  The hadronization of an expanding partonic fireball  is
studied within the Parton-Hadron-Strings Dynamics (PHSD) approach
which is based on a dynamical quasiparticle model (DQPM) matched
to reproduce lattice QCD results in thermodynamic equilibrium.
Apart from strong parton interactions the expansion and
development of collective flow is driven by strong gradients in the
parton mean-fields. An analysis of the elliptic flow
$v_2$ demonstrates a linear correlation with the spatial
eccentricity $\epsilon$ as in case of ideal hydrodynamics. The
hadronization occurs  by quark-antiquark fusion or 3 quark/3
antiquark recombination which is described by  covariant transition rates.
Since the dynamical quarks become very massive, the formed resonant 'pre-hadronic'
color-dipole states ($q\bar{q}$ or $qqq$) are of high invariant mass, too, and
sequentially decay to the groundstate meson and baryon octets increasing the total
entropy. This solves the entropy problem in hadronization in a
natural way. The resulting
particle ratios turn out to be in line with those from a grandcanonical
partition function at temperature $T \approx 170$ MeV rather
independent from the initial temperature and indicate an approximate
strangeness equilibration.
 \end{abstract}

\maketitle

\section{Introduction}

The 'Big Bang' scenario implies that in the first micro-seconds of the
universe the entire state has emerged from a partonic system of quarks,
antiquarks and gluons -- a quark-gluon plasma (QGP) -- to color neutral
hadronic matter consisting of interacting hadronic states (and
resonances) in which the partonic degrees of freedom are confined. The
nature of confinement and the dynamics of this phase transition has
motivated a large community for several decades  and is still an
outstanding question of todays physics. Early concepts of the QGP were
guided by the idea of a weakly interacting system of partons which
might be described by perturbative QCD (pQCD). However, experimental
observations at the Relativistic Heavy Ion Collider (RHIC) indicated
that the new medium created in ultrarelativistic Au+Au collisions was
interacting more strongly than hadronic matter (cf.\ \cite{QM01} and
Refs.\ therein) and consequently this concept had to be given up.
Moreover, in line with theoretical studies in Refs.
\cite{Shuryak,Thoma,Andre} the medium showed phenomena of an almost
perfect liquid of partons \cite{STARS,Miklos3} as extracted from the
strong radial expansion and elliptic flow of hadrons \cite{STARS}.

The question about the  properties of this (nonperturbative) QGP liquid
is discussed controversially in the literature  and dynamical concepts
describing the formation of color neutral hadrons from partons are
scarce \cite{Dyn,Bleicher,Koal1,Koal2,AMPT,Rapp,Biro}. A fundamental
issue for hadronization models is the conservation of 4-momentum as
well as the entropy problem because by fusion/coalescence of massless
(or low constituent mass) partons to color neutral bound states of low
invariant mass (e.g. pions) the number of degrees of freedom and thus
the total entropy is reduced in the hadronization process
\cite{Koal1,Koal2,AMPT}. This problem - a violation of the second law
of thermodynamics  as well as of the conservation 4-momentum and flavor
currents - definitely needs a sound dynamical solution.

A consistent dynamical approach - valid also for strongly interacting
systems - can be formulated on the basis of Kadanoff-Baym (KB)
equations \cite{KBaym,Sascha1} or off-shell transport equations in
phase-space representation, respectively \cite{Juchem,Sascha1,Knoll1}.
In the KB theory the field quanta are described in terms of propagators
with complex selfenergies.  Whereas the real part of the selfenergies
can be related to mean-field potentials, the imaginary parts  provide
information about the lifetime and/or reaction rates of time-like
'particles' \cite{Andre}. Once the proper (complex) selfenergies of the
degrees of freedom are known the time evolution of the system is fully
governed  by off-shell transport equations (as described in Refs.
\cite{Juchem,Sascha1,Knoll1}).

The determination/extraction of complex selfenergies for the partonic
degrees of freedom has been performed in Refs.
\cite{Andre,Cassing06,Cassing07} by fitting lattice QCD (lQCD) 'data'
within  the Dynamical QuasiParticle Model (DQPM). In fact, the DQPM
allows for a simple and transparent interpretation of lattice QCD
results for thermodynamic quantities as well as correlators and leads
to effective strongly interacting partonic quasiparticles with broad
spectral functions. We stress that mean-field potentials for the
'quarks' and 'gluons' as well as effective interactions have been
extracted from lQCD within the DQPM as well (cf. Ref.
\cite{Cassing07}).

\section{The PHSD approach}

The Parton-Hadron-String-Dynamics (PHSD) approach is a microscopic
covariant transport model that incorporates effective partonic as well
as hadronic degrees of freedom and involves a dynamical description of
the hadronization process from partonic to hadronic matter. Whereas the
hadronic part is essentially equivalent to the conventional
Hadron-Strings-Dynamics (HSD) approach \cite{HSD} the partonic dynamics
is based on the Dynamical QuasiParticle Model (DQPM)
\cite{Cassing06,Cassing07} which describes QCD properties in terms of
single-particle Green's functions (in the sense of a two-particle
irredicible (2PI) approach).

\subsection{Reminder of the DQPM}

We briefly recall the basic assumptions of the DQPM: Following Ref.
\cite{Andre05} the dynamical quasiparticle mass (for gluons and quarks) is
assumed to be given by the thermal mass in the asymptotic
high-momentum regime, which is proportional to the coupling
$g(T/T_c)$ and the temperature  $T$ with a running coupling
(squared),
\begin{eqnarray}
 g^2(T/T_c) = \frac{48\pi^2}{(11N_c - 2 N_f)
\ln(\lambda^2(T/T_c-T_s/T_c)^2}\ .
 \label{eq:g2}
\end{eqnarray}
Here $N_c = 3$ stands for the number of colors while $N_f$ denotes the
number of flavors. The parameters controlling the infrared enhancement
of the coupling $\lambda = 2.42$ and $T_s = 0.46 T_c$ have been fitted
in \cite{Andre05} to lQCD results for the entropy density $s(T)$. An
almost perfect reproduction of the energy density $\varepsilon(T)$ and
the pressure $P(T)$ from lQCD is achieved as well. As demonstrated in
Fig. 1 of Ref. \cite{Cassing07} this functional form for the strong
coupling $\alpha_s = g^2/(4\pi)$ is in full accordance with the lQCD
calculations of Ref. \cite{Bielefeld} for the long range part of the $q
- \bar{q}$ potential, too.

The width  for gluons and quarks (for vanishing chemical potential
$\mu_q$) is adopted in the form \cite{Pisar89LebedS}
\begin{eqnarray} \label{eq:gamma}
  \gamma_g(T)
  =
  \frac{3 g^2 T}{8 \pi} \,  \ln\left( \frac{2c}{g^2}\right)  \, , \
    \gamma_q(T)
  =
   \frac{g^2 T}{6 \pi} \,  \ln
  \left(\frac{2c}{g^2}\right)
  \, ,
\end{eqnarray}
where $c=14.4$ (from Ref. \cite{Andre}) is related to a magnetic
cut-off. We stress that a non-vanishing width $\gamma$ is the central
difference of the DQPM to conventional quasiparticle models
\cite{qp1,qp2,qp3}. It influence is essentially seen in correlation
functions as e.g. in the stationary limit of the correlation function
in the off-diagonal elements of the energy-momentum tensor $T^{kl}$
which defines the shear viscosity $\eta$ of the medium
\cite{Andre,Kubo}. Here a sizable width is mandatory to obtain a small
ratio in the shear viscosity to entropy density $\eta/s$.

In line with \cite{Andre05}
the parton spectral functions thus are no longer $\delta-$ functions in the
invariant mass squared but taken as
\begin{eqnarray}
 \rho_j(\omega)
 =
 \frac{\gamma_j}{ E_j} \l(
   \frac{1}{(\omega-E_j)^2+\gamma_j^2} - \frac{1}{(\omega+E_j)^2+\gamma_j^2}
 \r)
 \label{eq:rho}
\end{eqnarray} separately for quarks and gluons ($j=q,\bar{q},g$).
With the convention $E^2(\bm p) = \bm p^2+M_j^2-\gamma_j^2$, the
parameters $M_j^2$ and $\gamma_j$ are directly related to the real
and imaginary parts of the  retarded self-energy, e.g. $\Pi_j =
M_j^2-2i\gamma_j\omega$.

With the spectral functions fixed by Eqs.  (1)-(3) the total energy
density in the DQPM (at vanishing quark chemical potential) can be
evaluated as
\begin{equation} \label{ener}
T^{00} = d_g \int_0^\infty  \frac{d\omega}{2 \pi}
\int \frac{d^3 p}{(2 \pi)^3}\ 2 \omega^2 \rho_g(\omega, {\bf p})
n_B(\omega/T)
\end{equation}
$$ + d_q \int_0^\infty  \frac{d\omega}{2 \pi}
\int \frac{d^3 p}{(2 \pi)^3} \ 2 \omega^2 \rho_q(\omega, {\bf p})
n_F(\omega/T)$$ where $n_B$ and $n_F$ denote the Bose and Fermi
functions, respectively. The number of transverse gluonic degrees
of freedom is $d_g=16$ while the fermic degrees of freedom amount
to $d_q=4 N_c N_f=36$ in case of three flavors ($N_f$=3).
The pressure $P$ then may be obtained by integrating the
differential thermodynamic relation \begin{equation} P -
T \frac{\partial P}{\partial T}= - T^{00} \end{equation} with the
entropy density $s$ given by
\begin{equation}
s = \frac{\partial P}{\partial T} = \frac{T^{00}+P}{T} \ .
\end{equation}
This approach is thermodynamically consistent and represents a 2PI
approximation to hot QCD (once the free parameters in (1) and (2) are
fitted to lattice QCD results as in Refs.
\cite{Andre,Cassing06,Cassing07}).

As outlined in detail in Refs. \cite{Cassing06,Cassing07} the energy
density functional (\ref{ener}) can be separated in space-like and
time-like sectors when the spectral functions aquire a finite width.
The space-like part of (\ref{ener}) defines a potential energy density
$V_p$ since the field quanta involved are virtuell and correspond to
partons exchanged in interaction diagrams. The time-like part of
(\ref{ener}) corresponds to effective field quanta which can be
propagated within the light-cone. Related separations can be made for
virtuell and time-like parton densities \cite{Cassing06,Cassing07}.
Without repeating the details we mention that mean-field potentials for
partons can be defined by the derivative of the potential energy
density $V_p$ with respect to the time-like parton densities and
effective interactions by second derivatives of $V_p$ (cf. Section 3 in
Ref. \cite{Cassing07}).

\subsection{Hadronization in PHSD}

Based on the DQPM we have developed an off-shell transport approach
denoted as PHSD where the degrees-of-freedom are dynamical quarks,
antiquarks and gluons ($q, \bar{q}, g$) with rather large masses and
broad spectral functions in line with (1) - (3) as well as the
conventional hadrons (described in the standard HSD approach
\cite{HSD}). On the partonic side the following elastic and inelastic
interactions are included $qq \leftrightarrow qq$, $\bar{q} \bar{q}
\leftrightarrow \bar{q}\bar{q}$, $gg \leftrightarrow gg$, $gg
\leftrightarrow g$, $q\bar{q} \leftrightarrow g$  exploiting
'detailed-balance' with interaction rates from the DQPM
\cite{Andre,Cassing06,Cassing07} (cf. Section IIIB). The hadronisation,
i.e. transition from partonic to hadronic degrees of freedom, is
described by local covariant transition rates e.g. for $q+\bar{q}$
fusion to a meson $m$ of four-momentum $p= (\omega, {\bf p})$ at space
time point $x=(t,{\bf x})$:
\begin{eqnarray}
&&\phantom{a}\hspace*{-5mm} \frac{d N_m(x,p)}{d^4x d^4p}= Tr_q Tr_{\bar q} \
  \delta^4(p-p_q-p_{\bar q}) \
  \delta^4\left(\frac{x_q+x_{\bar q}}{2}-x\right) \nonumber\\
&& \times \omega_q \ \rho_{q}(p_q)
   \  \omega_{\bar q} \ \rho_{{\bar q}}(p_{\bar q})
   \ |v_{q\bar{q}}|^2 \ W_m(x_q-x_{\bar q},p_q-p_{\bar q}) \nonumber \\
&& \times N_q(x_q, p_q) \
  N_{\bar q}(x_{\bar q},p_{\bar q}) \ \delta({\rm flavor},\, {\rm color}).
\label{trans}
\end{eqnarray}
In (\ref{trans}) we have introduced the shorthand notation
\begin{equation}
Tr_j = \sum_j \int d^4x_j d^4p_j/(2\pi)^4
\end{equation}
where $\sum_j$ denotes a summation over discrete quantum numbers
(spin, flavor, color); $N_j(x,p)$ is the phase-space density of
parton $j$ at space-time position $x$ and four-momentum $p$.  In
Eq. (\ref{trans}) $\delta({\rm flavor},\, {\rm color})$ stands
symbolically for the conservation of flavor quantum numbers as
well as color neutrality of the formed hadron $m$ which can be
viewed as a color-dipole or 'pre-hadron'.  Furthermore, $v_{q{\bar
q}}(\rho_p)$ is the effective quark-antiquark interaction  from
the DQPM  defined by Eq. (31) and displayed in Fig. 10 of Ref.
\cite{Cassing07} as a function of the local parton ($q + \bar{q}
+g$) density $\rho_p$ (or energy density). Furthermore, $W_m(x,p)$
is the phase-space distribution of the formed 'pre-hadron'. It is
taken as a Gaussian in coordinate and momentum space (following
Ref. \cite{Dover}) with width $\sqrt{<r^2>}$ = 0.66 fm (in the rest
frame) which corresponds to an average rms radius of mesons.  The
width in momentum space is fixed by the  uncertainty
principle, i.e. $\Delta x \Delta p = 1$ (in natural units). We
note that the final hadron formation rates are approximately
independent on these parameters within reasonable variations.

In principle,  the two-particle Greens-function
$G^<(x_q,p_q,x_{\bar q},p_{\bar q})$ should appear  in Eq.
(\ref{trans}).  The approximation of the two-particle Greens
function by a (symmetrized/antisymmetrized) product of
single-particle Greens functions is always a first step in a
cluster expansion for Greens functions and neglects 'residual
correlations' stemming from higher order contractions. The same
holds for an approximation of the three-particle Greens function
by the (symmetrized/antisymmetrized) product of single-particle
Greens functions (cf. Ref. \cite{Wang1}). However, the DQPM with its
dynamical spectral functions already includes the effects of
strong two-body correlations - contrary to bare Green functions -
such that the effect of residual interactions might be discarded
in a first approximation. But there is no 'a priori' guarantee
that this approximation is appropriate under all circumstances.
This, in principle, should be examined by lattice QCD in order to
test the cluster decomposition in hot QCD.

Related transition rates (to Eq. (\ref{trans})) are defined for the
fusion of three off-shell quarks ($q_1+q_2+q_3 \leftrightarrow B$)
to a color neutral baryonic ($B$ or $\bar{B}$) resonances of
finite width (or strings) fulfilling energy and momentum
conservation as well as flavor current conservation, i.e.
\begin{eqnarray}
&&\phantom{a}\hspace*{-5mm}  \frac{d N_B(x,p)}{d^4x d^4p}=
Tr_{q_1} Tr_{q_2} Tr_{q_3} \  \delta^4(p-p_1-p_2-p_3) \ \nonumber\\
&&  \delta^4\left( \frac{x_{q_1}+x_{q_2}+x_{q_3}}{3}-x \right)
              \label{trans2}\\
&& \times \omega_{q_1} \ \rho_{q_1}(p_1)
   \  \omega_{q_2} \ \rho_{{q_2}}(p_2)  \  \omega_{q_3} \ \rho_{{q_3}}(p_3)
       \nonumber\\
&& \times |M_{qqq}|^2 \ W_B(x_1,x_2,x_3,p_1,p_2,p_3) \nonumber \\
&& \times N_{q_1}(x_1, p_1) \  N_{q_2}(x_2,p_2) \
  N_{q_3}(x_3,p_3) \ \delta({\rm flavor, color}). \nonumber
\end{eqnarray}
Here the quantity $W_B$ denotes the baryon phase-space
distribution (evaluated in Jacobi coordinates) which is taken
again of Gaussian shape with a width of 1 fm in coordinate space
which corresponds to an average rms radius of excited baryons. The
matrix element squared $|M_{qqq}|^2$ reflects the strength of
three quark fusion processes and is fixed as follows: Since Regge
trajectories for excited mesonic and baryonic states have the same
slope (or string constant in the color dipole picture) we
tentatively set $|M_{qqq}|^2 = |v_{q\bar{q}}|^3$ in our present
work which implies that (so far) there is no need to introduce any new
parameter.

On the hadronic side PHSD includes explicitly the  baryon octet and
decouplet, the $0^-$ and $1^-$ meson nonets as well as selected higher
resonances as in HSD \cite{HSD}. Hadrons of higher masses ($>$ 1.5 GeV
in case of baryons and $>$ 1.3 GeV in case of mesons) are treated as
'strings' (color-dipoles) that  decay to the known (low mass) hadrons
 according to the JETSET algorithm \cite{JETSET}.

\section{Hadronization of an expanding partonic fireball}

We now turn to actual results from PHSD for the model case of an
expanding partonic fireball at initial temperature $T=1.7\  T_c$
($T_c$= 0.185 GeV) with quasiparticle properties and four-momentum
distributions determined by the DQPM at temperature $T$ = 1.7 $T_c$.

\subsection{Initial conditions}
The initial distribution for quarks, antiquarks and gluons
in coordinate space
is taken as a Gaussian ellipsoid with a spatial eccentricity
\begin{equation}
\epsilon =\langle y^2-x^2\rangle/\langle y^2 + x^2\rangle
\end{equation}
and $<z^2> = <y^2>$ in order to allow for the built-up of elliptic flow
(as in semi-central nucleus-nucleus collisions at relativistic
energies). In order to match the initial off-equilibrium strange quark
content in relativistic $pp$ collisions the number of $s$ (and
$\bar{s}$ quarks) is assumed to be suppressed by a factor of 3 relative
to the abundance of $u$ and $d$ quarks and antiquarks. In this way we
will be able to investigate additionally the question of strangeness
equilibration.

As mentioned above the dynamical evolution of the system is then
entirely described by the transport dynamics in PHSD incorporating
the off-shell propagation of the partonic quasiparticles according
to Refs. \cite{Juchem} as well as the transition to resonant
hadronic states (or 'strings') Eqs. (\ref{trans}), (\ref{trans2}).
The time integration for the testparticle equations of motion (cf.
Refs. \cite{Juchem}) is performed in the same way as in case of
hadronic off-shell transport where (in view of the
momentum-independent width $\gamma$) the simple relation (19) in
Ref. \cite{NPA807} is employed. For the collisions of partons
two variants are at our disposal: i) geometrical collision
criteria as employed in standard hadronic transport, ii) the
in-cell method developed in Ref. \cite{Lang}. The latter can
easily be extended to describe $2 \leftrightarrow 3$ processes
etc. in a covariant way \cite{NPA700} and is the better choice at
high particle densities (cf. Ref. \cite{XU}).  The hadronization
is performed by integrating the rate equations (\ref{trans}) and
(\ref{trans2}) in space and time which are discretized by $\Delta
t$ and $\Delta V(t)$. We use local cells of volume $dV(t)$= 0.25
(1+$b t)^3$ fm$^3$ where $t$ is given in units of fm/c and $b$=
0.7 c/fm. This choice approximately corresponds to a comoving grid
for the expanding system. In each time-step $\Delta t$ and cell
$\Delta V$ the integrals in (\ref{trans}) and (\ref{trans2}) are
evaluated by a sum over all (time-like) testparticles using
(e.g. for the quark density)
\begin{eqnarray}
&&\frac{1}{\Delta V} \int_{\Delta V} d^3x  \int \frac{d \omega_q}{2
\pi} \omega_q  \int \frac{d^3  p_q}{(2\pi)^3} \
\rho_q(\omega_q,p_q)\ N_q(x,p_q) \nonumber\\
&&= \frac{1}{\Delta V} \sum_{J_q \varepsilon \Delta V}  1 \
=  \ \rho_q(\Delta V) \ ,
\label{rho_DV}
\end{eqnarray}
where the sum over $J_q$ implies a sum over all
testparticles of type $q$ (here quarks) in the local volume
$\Delta V$ in each parallel run. In order to obtain lower
fluctuations the integrals are averaged over the number of
parallel runs (typically a few hundred). For each individual
testparticle (i.e. $x_q$ and $p_q$ fixed) the additional
integrations in (\ref{trans}) and (\ref{trans2}) give a
probability for a hadronization process to happen; the actual
event then is selected by Monte Carlo. Since energy-momentum
conservation fixes the four-momentum $p$ of the hadron produced -
the space-time position $x$ is fixed by (7) or (9) -  the latter
is represented by a hadronic state with flavor content fixed by
the fusing quarks (antiquarks). The latter decays to the lower
mass hadrons according to JETSET \cite{JETSET} above thresholds of
1.3 GeV for mesonic states or 1.5 GeV for baryonic states (as in
HSD).  Lower mass hadrons (octet and decouplet states) are
determined by the weight of their respective spectral functions at
given invariant mass and selected by Monte Carlo. Note that the
propagation of partons includes the space-time derivatives of the
quark and gluon mean-fields specified in Eq. (29)  and displayed
in Fig. 9 of Ref. \cite{Cassing07}.

\subsection{Dynamical evolution}
\begin{figure}[t]
\centering \includegraphics*[width=85mm]{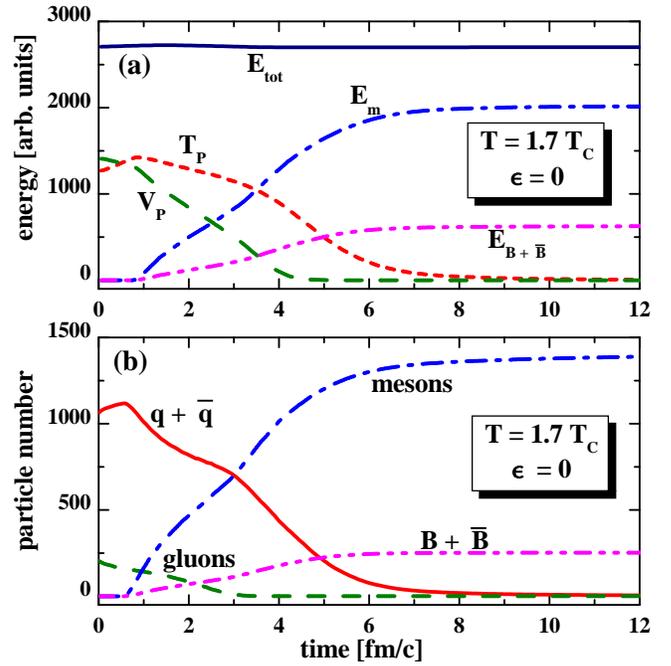}
\caption{(Color online) Upper part: Time evolution of the total energy
$E_{tot}$ (upper line), the partonic contributions from the interaction
energy $V_p$ and the energy of time-like partons $T_p$ in comparison to
the energy contribution from formed mesons $E_m$ and baryons (+
antibaryons) $E_{B+{\bar B}}$.  Lower part:  Time evolution in the
parton, meson and baryon number for an expanding partonic fireball at
initial temperature $T=1.7\  T_c$ with initial eccentricity $\epsilon = 0$.}
\label{fig1}
\end{figure}

In Fig. \ref{fig1} (upper part) we show the energy balance for the
expanding system at initial temperature $T= 1.7 T_c$ and
eccentricity $\epsilon$ =0, i.e. a fireball of spherical symmetry.
The total energy $E_{tot}$ (upper line) - which at $t=0$ is given
by (\ref{ener}) integrated over space - is conserved within 3\%
throughout the partonic expansion and hadronization phase such
that for $t > 8$ fm/c it is given essentially by the energy
contribution from mesons and baryons (+antibaryons). The initial
energy splits into the partonic interaction energy $V_p$ (cf. Eq.
(19) in Ref. \cite{Cassing07}) and the energy of the time-like
(propagating) partons
\begin{equation} \label{TK}
T_p = \sum_i \sqrt{p^2_i + M^2_i(\rho_p)}
\end{equation}
with fractions determined by the DQPM \cite{Cassing07}. In Eq.
(\ref{TK}) the summation over $i$ runs over all testparticles in an
individual run. The hadronization mainly proceeds during the time
interval 1 fm/c $< t < $ 7 fm/c (cf.  the lower part of Fig. \ref{fig1}
where the time evolution of the $q, \bar{q},g$, meson and baryon
(+antibaryon) number is displayed).

As one observes from Fig. \ref{fig1}  on average the number of hadrons
from the resonance or 'string' decays is larger than the initial number
of fusing partons.  This might be astonishing since by partonic fusion
the number of final states is conventionally reduced in coalescence
models.

\begin{figure}[t]
\centering \includegraphics*[width=85mm]{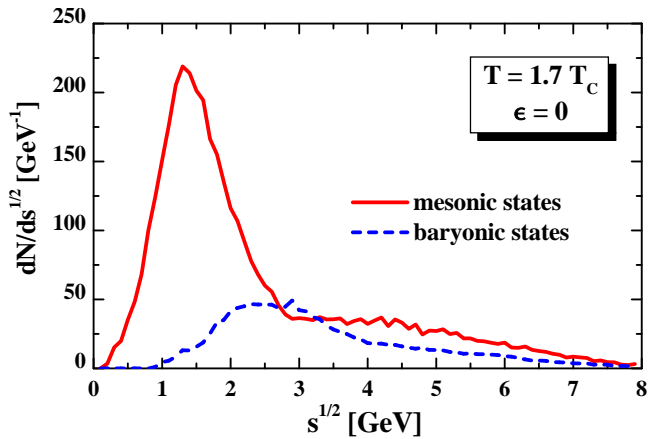}
\caption{(Color online) The invariant mass distribution for fusing $q
\bar{q}$ pairs (solid line) as well as $qqq$ (and
$\bar{q}\bar{q}\bar{q}$) triples (dashed line) that lead to the
formation of final hadronic states for an expanding partonic fireball
at initial temperature $T=1.7\  T_c$ with initial eccentricity
$\epsilon = 0$.} \label{fig2a}
\end{figure}

In order to shed some light on the hadronization process in PHSD we
display in Fig. \ref{fig2a} the invariant mass distribution of $q
\bar{q}$ pairs (solid line) as well as $qqq$ (and
$\bar{q}\bar{q}\bar{q}$) triples (dashed line) that lead to the
formation of final hadronic states. In fact, the distribution for the
formation of baryon (antibaryon) states starts above the nucleon mass
and extends to  high invariant mass covering the nucleon resonance mass
region as well as the high mass continuum (which is treated by the
decay of strings within the JETSET model \cite{JETSET}). On the
'pre-mesonic side the invariant-mass distribution starts roughly above
the two-pion mass and extends up to continuum states of high invariant
mass (described again in terms of string excitations). The low mass
sector is dominated by $\rho$, $a_1$, $\omega$ or $K^*, \bar{K}^*$
transitions etc. The excited 'pre-hadronic' states  decay to two or
more 'pseudoscalar octet' mesons such that the number of final hadrons
is larger than the initial number of fusing partons.

Accordingly,  the hadronization process in PHSD leads to an increase of
the total entropy and not to a decrease as in case of coalescence
models \cite{Koal1,Koal2}. This is a direct consequence of the finite
(and rather large) dynamical quark and antiquark masses as well as
mean-field potentials which - by energy conservation - lead to
'pre-hadron' masses well above those for the pseudo-scalar meson octett
or the baryon octett, respectively. This solves the entropy problem in
hadronization in a natural way and is in accordance with the second law
of thermodynamics!

\begin{figure}[t]
\centering \includegraphics*[width=85mm]{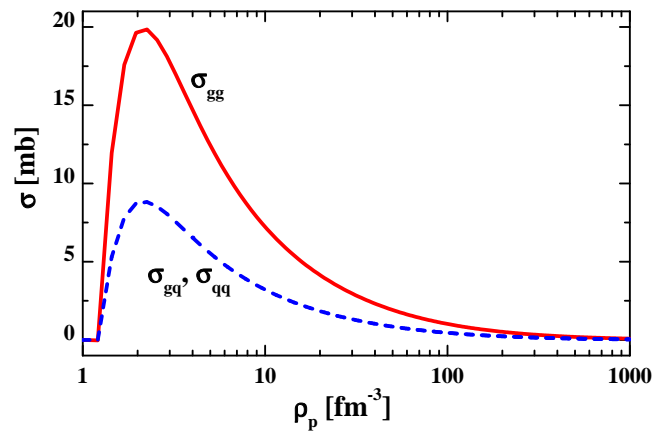}
\caption{(Color online) The effective gluon-gluon $\rightarrow$
gluon-gluon (solid line), gluon-quark $\rightarrow$ gluon-quark and
quark-quark $\rightarrow$ quark-quark (dashed line) cross section from
the DQPM as a function of the parton density $\rho_p$. }
\label{fig3an}
\end{figure}

\begin{figure}[t]
\centering \includegraphics*[width=85mm]{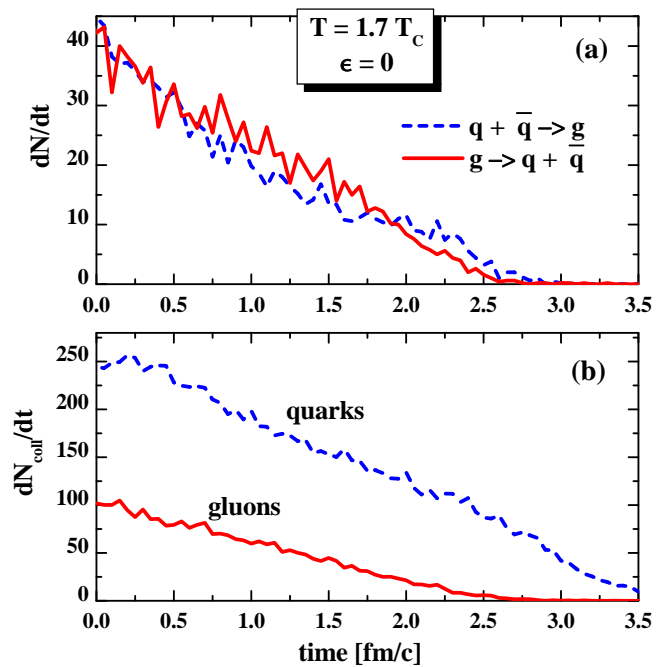}
\caption{(Color online) The interaction rate for the channels
$q+\bar{q} \rightarrow g$ (dashed line, upper part) and $g \rightarrow
q+\bar{q}$ (red line, upper part) for an expanding partonic fireball at
initial temperature $T=1.7\ T_c$ with initial eccentricity $\epsilon = 0$.
The lower part shows the collision rate of gluons (solid line) and
quarks or antiquarks (dashed line). }
\label{fig3n}
\end{figure}

The parton dynamics itself is governed by their propagation in the
time-dependent mean-field $U_p(\rho_p)$ which is adopted in the
parametrized form (as a function of the parton density $\rho_p$) given
by Eq. (29) in Ref. \cite{Cassing07}. Since the mean-field $U_p$ is
repulsive the partons are accelerated during the expansion phase on
expense of the potential energy density $V_p$ which is given by the
integral of $U_p$ over $\rho_p$ (cf. Section 3 in Ref.
\cite{Cassing07}). The interaction rates of the partons are determined
by effective cross sections which for  $gg$ scattering have been
determined in Ref. \cite{Andre} as a function of $T/T_c$. The latter
are re-parametrized in the actual calculation as a function of the
parton density using the available dependence of $\rho_p(T)$ on the
temperature $T$ from the DQPM.

The actual values for $gg$ scattering are shown in Fig.
\ref{fig3an} as a function of the parton density $\rho_p$ (solid
line) and demonstrate that $gg$  cross sections up to 20 mb can be
reached at $\rho_p \approx 2.5$ fm$^{-3}$. The effective cross
section drops rapidly with increasing $\rho_p$ which signals that
weakly interacting partons might show up at very high parton
density. Some note of caution has to be added here since though
the cross section $\sigma_{gg}$ drops with $\rho_p$ the collision
rate of a gluon ($\sim \sigma_{gg} \rho_p$) increases slightly
with $\rho_p$. For quark-quark or quark-antiquark elastic
scattering the cross section is reduced by a factor 4/9 in line
with Casimir scaling. Quark-gluon elastic scattering (in the
present implementation) is also reduced by a factor 4/9 (cf. Fig.
\ref{fig3an}, dashed line).  The channels $q \bar{q} \rightarrow g
$ are described by a relativistic Breit-Wigner cross section which
is determined by the actual masses of the fermions, the invariant
energy $\sqrt{s}$ and the resonance parameters of the gluon (from
the DQPM). In this case a further constraint on flavor neutrality
and open color is employed. The gluon decay to a $u \bar{u}, d
\bar{d}$ or $s\bar{s}$ pair is fixed by detailed balance. Further
channels are $gg \leftrightarrow g$ which are given by
Breit-Wigner cross sections (with the gluon resonance parameters)
and detailed balance, respectively.

\begin{figure}[t]
\centering \includegraphics*[width=85mm]{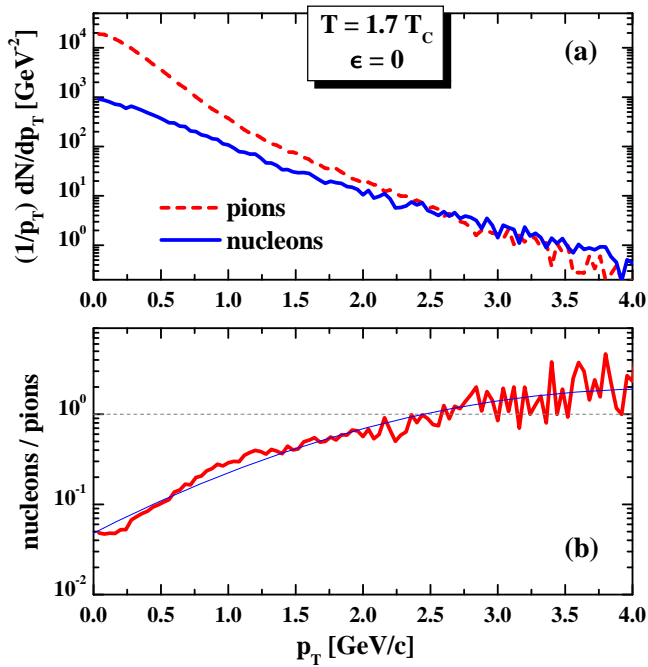}
\caption{(Color online) Upper
part: The transverse momentum spectra $1/p_T dN/dp_T$ for pions
and nucleons for an expanding partonic fireball at initial
temperature $T=1.7\  T_c$ with initial eccentricity $\epsilon =
0$.  Lower part: The nucleon to pion ratio as a function of the
transverse momentum $p_T$ corresponding to the spectra in the
upper part. The thin blue line is drawn to guide the eye due to
the limited statistics of the PHSD calculation.} \label{fig2b}
\end{figure}

The actual interaction rates for the channels $q+\bar{q} \rightarrow g$
and $g \rightarrow q+\bar{q}$ are displayed in Fig. \ref{fig3n} by the
dashed line (upper part) and the solid line (upper part), respectively,
for the expanding partonic fireball at initial temperature $T=1.7\ T_c$
with initial eccentricity $\epsilon = 0$.  Within statistics the
numerical result shows that detailed balance actually is fulfilled for
the expanding partonic system which was initialized in thermal
equilibrium. The lower part of Fig.  \ref{fig3n} shows the total number
of collisions per time for gluons (solid line) and quarks or antiquarks
(dashed  line) which is higher for the fermions since the latter are
much more frequent than the gluons.

Let's, furthermore, have a look at the transverse momentum $p_T$
spectra of hadrons emerging from the PHSD dynamics of the expanding
fireball initialized with $T= 1.7 \ T_c$. The resulting $p_T$ spectra
are displayed in Fig. \ref{fig2b} (upper part) for pions and nucleons
and show that the nucleons become more abundant for $p_T >$ 2.5 GeV/c.
The ratio of nucleons to pions is depicted in the lower part of Fig.
\ref{fig2b} and clearly demonstrate that baryons (antibaryons) become
more frequent than mesons at high $p_T >$ 2.5 GeV. This observation is
in close analogy to the experimental findings in Au+Au collisions at
top RHIC energies \cite{STARS}. Note that a quantitative comparison
with RHIC data is not meaningful due to the rather simplified and
special initial conditions employed here.

\subsection{Comparison to the statistical hadronization model}

It is, furthermore, interesting to have a look at the final particle
ratios $K^+/\pi^+$, $p/\pi^+$, $\Lambda/K^+$ etc. (after hadronic
decays) which are shown in Table 1. The latter ratios are compared to
the grandcanonical statistical hadronization model (SM)
\cite{PBM,PBM2,Anton} at baryon chemical potential $\mu_B = 0$.  For
$\mu_B = 0$ the particle ratios depend only on temperature $T$ and one
may fix a freeze-out temperature, e.g., by the proton to $\pi^+$ ratio.
A respective comparison is given also in Table 1 for $T$ = 160 MeV and
170 MeV for the SM which demonstrates that the results from PHSD are
close to those from the SM for $T \approx$ 170 MeV. This also holds
roughly for the $\Lambda/K^+$ ratio. On the other hand the $K^+/\pi^+$
ratio only smoothly depends on the temperature $T$ and measures the
amount of strangeness equilibration. Recall that we initialized the
system with a relative strangeness suppression factor of 1/3. The
deviation from the SM ratio by about 13\% indicates that strangeness
equilibration is not fully achieved in the calculations. This is
expected since the partons in the surface of the fireball hadronize
before chemical equilibration may occur. A detailed discussion of
results will be presented in a forthcoming study.

\begin{table}[h]
\begin{center}
\begin{tabular}{|c|c|c|c|} \hline
\hspace*{3.0cm} &~~~~$p/\pi^+$~~~  & ~~~$\Lambda/K^+$~~~   & ~~~$K^+/\pi^+$~~~~   \\
 \hline
PHSD  &    0.086    & 0.28            &   0.157         \\ \hline
SM $T =160$~MeV &   0.073   & 0.22   &  0.179 \\          \hline
SM $T =170$~MeV &   0.086  & 0.26   &  0.180 \\
 \hline
  \end{tabular}
\end{center}
\caption{Comparison of particle ratios from PHSD with the
statistical model (SM) \cite{Anton} for $T$= 160 MeV and 170 MeV.}
\end{table}

The agreement between the PHSD and SM results for the baryon to meson
ratio in the strangeness S=0 and S=1 sector may be explained as
follows:  Since the final hadron formation dominantly proceeds via
resonance and string formation and decay - which is approximately
a microcanonical statistical process \cite{beccatini} - the
average over many hadronization events with different energy/mass
and particle number (in the initial and final state) leads to a
grandcanonical ensemble. The latter (for $\mu_B = 0$) is only
characterized by the average energy or an associated Lagrange
parameter $\beta=1/T$.

\subsection{Elliptic flow}
\begin{figure}[t]
\centering \includegraphics*[width=85mm]{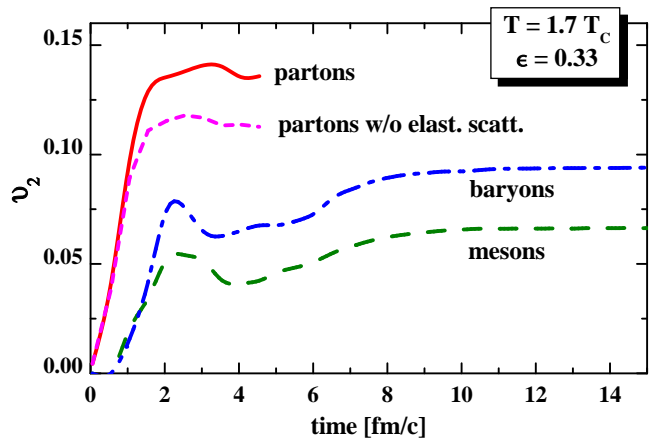}
\caption{(Color online) Time evolution of the elliptic flow $v_2$
for partons  and hadrons for the initial spatial eccentricity
$\epsilon =0.33$ for an expanding partonic fireball at initial temperature
$T=1.7\  T_c$.}
 \label{fig2}
\end{figure}

Of additional interest are the collective properties of the
partonic system during the early time evolution. In order to
demonstrate the built up of elliptic flow we show in Fig.
\ref{fig2} the time evolution of \begin{equation} v_2 =
\left\langle (p_x^2 - p_y^2)/(p_x^2 + p_y^2) \right\rangle
\end{equation} for partons (solid line), mesons (long dashed line)
and baryons (dot-dashed line) for an initial eccentricity
$\epsilon = 0.33$. As seen from Fig. \ref{fig2} the partonic flow
develops within 2 fm/c and the hadrons  essentially pick up the
collective flow from the accelerated partons. The hadron $v_2$ is
smaller than the maximal parton $v_2$ since by parton fusion the
average $v_2$ reduces and a fraction of hadrons is formed early at
the surface of the fireball without a strong acceleration before
hadronization. We briefly mention that the reduction of the
average hadron elliptic flow essentially is due to the finite
parton masses which are larger than the temperature in the
hadronization phase. This reduced hadron $v_2$ is in contrast to
the coalescence of massless partons. We discard a more detailed
discussion and investigation of Eqs. (\ref{trans}) and
(\ref{trans2}) and refer the reader to an upcoming study.

It is important to point out that in PHSD the elliptic flow of
partons predominantly stems from the gradients of the repulsive
parton mean-fields (from the DQPM) at high parton (energy)
density. To demonstrate this statement we show in Fig. \ref{fig2}
the result of a simulation without elastic partonic rescattering
processes by the short dashed line.

\begin{figure}[t]
\centering \includegraphics*[width=85mm]{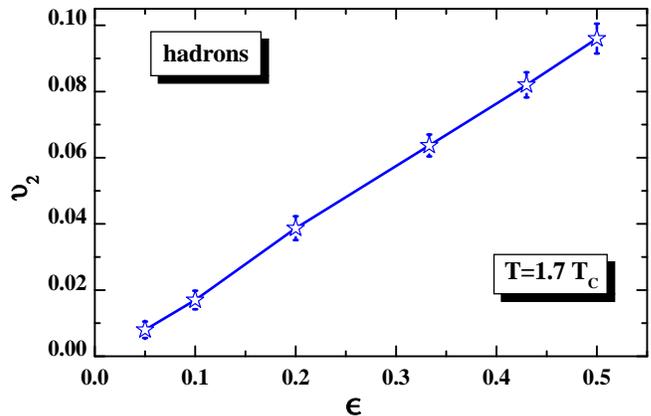}
\caption{(Color online) The hadron elliptic flow $v_2$
versus the initial spatial eccentricity
$\epsilon$  for an expanding partonic fireball at initial
temperature $T=1.7\  T_c$.} \label{fig3}
\end{figure}

Fig. \ref{fig3} shows the final hadron $v_2$ versus the initial
eccentricity $\epsilon$ and indicates that the ratio
$v_2/\epsilon$ is practically constant ($\approx 0.2$) as in ideal
hydrodynamics (cf. Fig. 3 in Ref. \cite{Voloshin}). Accordingly
the parton dynamics in PHSD are close to ideal hydrodynamics. This
result is expected since the ratio of the shear viscosity $\eta$
to the entropy density $s$ in the DQPM is on the level of $\eta/s
\approx $ 0.2 \cite{Andre} and thus rather close to the lower
bound of $\eta/s = 1/(4 \pi)$ \cite{Son}. Note that the ratio
$\eta/s$ is dominantly determined by the quasiparticle width
$\gamma$ (\ref{eq:gamma}) and low ratios on the level of $\eta/s
\approx$  0.2 require broad parton spectral functions as employed
in the DQPM.

\begin{figure}[t]
\centering \includegraphics*[width=85mm]{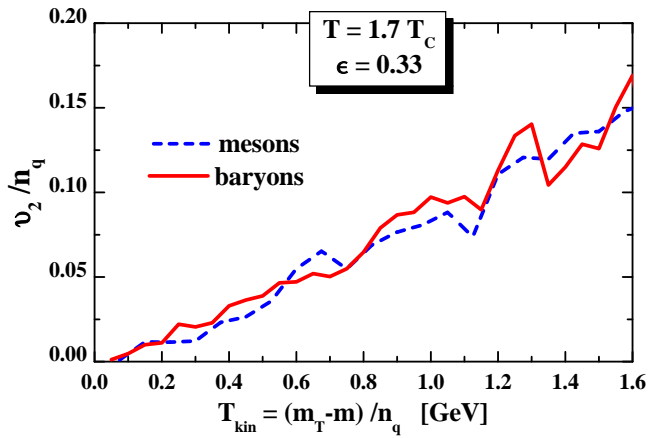} \caption{(Color online) The
elliptic flow $v_2$ - scaled by the number of constituent quarks
$n_q$ - versus the transverse kinetic energy divided by $n_q$ for
mesons (dashed line) and baryons (solid line). }
\label{fig5}
\end{figure}

A further test of the PHSD hadronization approach is provided by the
 'constituent quark number scaling' of the elliptic flow $v_2$ which
 has been observed experimentally
in central Au+Au collisions at RHIC \cite{STARS,Star08}. In this
respect we plot $v_2/n_q$ versus the transverse kinetic energy per
constituent parton,
\begin{equation} T_{kin} = \frac{m_T-m}{n_q} \ , \end{equation}
with $m_T$ and $m$  denoting the transverse mass and actual mass,
respectively. For mesons we have $n_q = 2$ and for
baryons/antibaryons $n_q=3$. The results for the scaled elliptic
flow are shown in Fig. \ref{fig5} for mesons (dashed  line) and
baryons (solid line) and suggest an approximate scaling. We note
that the scaled hadron elliptic flow $v_2/n_q$ does not reflect
the parton $v_2$ at hadronization and is significantly smaller.
Due to the limited statistics especially in the baryonic sector
with increasing $p_T$ this issue will have to be re-addressed with
high statistics in the actual heavy-ion case where the very early
parton $p_T$ distribution also shows 'power-law' tails.

\section{Summary}
In summary, the expansion dynamics of an anisotropic partonic
fireball is studied within the PHSD approach which includes
dynamical local transition rates from partons to hadrons
(\ref{trans}), (\ref{trans2}) (and vice versa). It shows collective
features as expected from ideal hydrodynamics in case of strongly
interacting systems. The hadronization process conserves
four-momentum and all flavor currents and slightly increases the
total entropy (by about 15\% in the model case investigated here)
since the 'fusion' of rather massive partons dominantly leads to
the formation of color neutral strings or resonances that decay
microcanonically to lower mass hadrons. This solves the entropy
problem associated with the simple coalescence model!

We find that the hadron abundancies and baryon to meson ratios are
compatible with those from the statistical hadronization model
\cite{PBM,PBM2} - which describes well particle ratios from AGS to
RHIC energies - at a freeze-out temperature of about 170 MeV.
Furthermore, strangeness equilibration is approximately achieved
in the dynamical expansion and driven by the processes $q\bar{q}
\leftrightarrow g \leftrightarrow s\bar{s}$, which is a resonant
process in the DQPM. However, although the final hadron ratios are
compatible with a fixed freeze-out temperature ($\sim$ 170 MeV) we
observe that the actual hadronization occurs at very different energy
densities (or temperatures) (cf. also Ref. \cite{Knoll08}) such that
the microscopic studies do not support the sudden freeze-out
picture.

Our calculations show that the hadron elliptic flow is essentially
produced in the early partonic stage where also the strong repulsive
parton mean-fields contribute to a large extent. This might explain why
the hadron $v_2$ from HSD calculations \cite{Brat03} underestimated the
RHIC data on $v_2$ essentially at midrapidity (in the pure
hadron/string approach). The hadron elliptic flow from PHSD is smaller
than the parton $v_2$ due to a partial cancellation of the $v_2$ values
from the individual partons in the fusion process to hadrons (but
larger than in HSD). This cancellation essentially is a consequence of
parton  masses that are larger than the 'local temperature' during the
hadronization phase. Nevertheless, our observations indicate an
approximate quark number scaling (cf. Fig. \ref{fig5}) for low and
moderate transverse kinetic energies within the statistics reached so
far.  This issue will be followed up in more detail in a forthcoming
investigation.

The present study, however, serves only as a model case which allows
for a more transparent interpretation of the various results. An
application of the PHSD approach to ultrarelativistic heavy-ion
collisions especially in comparison to differential experimental data
is expected to shed further light on the transport properties of the
partonic phase and the dynamics of hadronization.

\phantom{a}\vspace{0mm} The authors  like to thank A. Andronic for
providing the SM results in Table 1 and for a critical
reading of the manuscript. Furthermore, they are grateful to O. Linnyk
for valuable discussions.


\end{document}